# Flatband Line States in Photonic Super-Honeycomb Lattices

Wenchao Yan[1], Hua Zhong[2], Daohong Song[1,3*], Yiqi Zhang[2], Shiqi Xia[1], Liqin Tang[1,3], Daniel Leykam[4], and Zhigang Chen[1,3,5*]

[1]*The MOE Key Laboratory of Weak-Light Nonlinear Photonics, TEDA Applied Physics Institute and School of Physics, Nankai University, Tianjin 300457, China*
[2]*Department of Applied Physics, School of Science, Xi'an Jiaotong University, Xi'an 710049, China*
[3]*Collaborative Innovation Center of Extreme Optics, Shanxi University, Taiyuan, Shanxi 030006, People's Republic of China*
[4]*Center for Theoretical Physics of Complex Systems, Institute for Basic Science, Daejeon 34126, Republic of Korea*
[5]*Department of Physics and Astronomy, San Francisco State University, San Francisco, California 94132, USA*
*songdaohong@nankai.edu.cn, zgchen@nankai.edu.cn*

**Abstract:** We establish experimentally a photonic super-honeycomb lattice (sHCL) by use of a cw-laser writing technique, and thereby demonstrate two distinct flatband line states that manifest as noncontractible-loop-states in an infinite flatband lattice. These localized states ("straight" and "zigzag" lines) observed in the sHCL with tailored boundaries cannot be obtained by superposition of conventional compact localized states because they represent a new topological entity in flatband systems. In fact, the zigzag-line states, unique to the sHCL, are in contradistinction with those previously observed in the Kagome and Lieb lattices. Their momentum-space spectrum emerges in the high-order Brillouin zone where the flat band touches the dispersive bands, revealing the characteristic of topologically protected bandcrossing. Our experimental results are corroborated by numerical simulations based on the coupled mode theory. This work may provide insight to Dirac-like 2D materials beyond graphene.

Flatband systems, periodic lattices hosting at least one completely dispersionless energy band, have attracted enormous interest in different branches of physics ranging from condensed matter to exciton polaritons, ultracold atoms and optics[1-15]. In optics, photonic lattices composed of evanescent coupled waveguide arrays provide a promising platform to explore intriguing flatband phenomena associated with the lattice models originally proposed in solid state physics[16,17], many of which are difficult to be realized in electronic systems. Flatband lattices host localized states commonly referred to as compact localized states (CLSs)[1,2], typically localized over a few lattice unit cells, while their wavefunction amplitude vanishes on all other sites due to destructive interference[18,19]. The CLSs have been experimentally realized in various one-dimensional (1D) and two-dimensional (2D) flatband photonic lattices[20-25], and they have also been proposed for applications such as light localization, distortion-free imaging and even flatband lasing[1,2,26].

In certain flatband lattices hosting a band crossing between the flat band and other dispersive bands at discrete points in momentum space, the CLSs were found to form an incomplete basis for the flatband. The missing states are noncontractible loop states (NLSs) that wind around the entire (infinite) lattices, a real space manifestation of the Brillouin zone's torus topology[27,28]. The NLSs cannot be obtained by linear superposition of the conventional CLSs and form a new type of flatband eigenstates. Since an infinite lattice or a torus structure is difficult to establish in experiment, an alternative approach is to observe the "line states" in truncated lattices with appropriate boundaries. The existence of the line states does not rely on the spatial size of the lattice but rather on the boundary termination. Recently, such line states have been experimentally observed in a finite-sized photonic Lieb lattice with bearded edges[29], where a NLS manifests its existence as a straight line. For Lieb lattices, the flatband touching occurs at the corners of the first Brillouin zone (BZ). The flatband touching can also appear at other high symmetry points in momentum space such as the BZ centers, as in a Kagome lattice and a super-honeycomb lattice (sHCL).

The sHCL is particularly interesting because it is an example of a lattice where fermionic (spin-½) and bosonic (spin-1) conical band crossing points coexist. Such lattice structures were initially studied more than two decades ago in electronic systems[30,31], and were recently revisited by theorists studying their Dirac cones and conical diffraction[32-34]. One wonders if the composite sHCL could also be realized in experiment, and how its qualitatively different band structure would affect the NLSs.

In this work, by employing a cw-laser writing technique, we establish a finite-sized sHCL in the bulk of a nonlinear crystal. More importantly, we experimentally observe two types of line states (the "straight" and "zigzag" lines) in the sHCL with tailored boundaries which cannot be obtained by superposition of conventional flatband CLSs. The straight line states occupy two majority sublattices, while the zigzag line states reside in three majority sublattices. The latter states are unique for the sHCL, which have never been predicated or observed whatsoever. Our theoretical analysis shows that the zigzag line states are related to linear superposition of straight line states. Surprisingly, the *k*-space spectra of these line states emerge at the high-order BZ (across the centers of the second BZ) where the flat band touches the dispersive bands, in contrast to that of the line states in the Lieb lattices[28]. These results validate further the existence of the NLSs as a direct manifestation of band touching arising from real space topology, and they may prove relevant to 2D materials with flatbands and band-touching spectrum.

The sHCL, also known as edge-centered honeycomb lattice, consists of five lattice sites (A, B, C, D, E) per unit cell as marked in Fig. 1(a), which can be considered as introducing additional lattice sites C, D, E (majority) to the HCL sites A, B (minority). A light beam propagating in the 2D sHCL is governed by the Schrödinger-type paraxial wave equation:

$$i\frac{\partial \Psi(x,y,z)}{\partial z} = -\frac{1}{2k_0}\nabla^2 \Psi(x,y,z) - \frac{k_0 \Delta n(x,y)}{n_0}\Psi(x,y,z) \equiv H_0 \Psi(x,y,z). \quad (1)$$

Here $\Psi$ is the electric field envelope of the probe beam, $\nabla^2 = \partial_x^2 + \partial_y^2$ is the transverse Laplacian operator, $z$ is the longitudinal propagation distance into the photonic lattice, $k_0$ is the wavenumber in the medium, $n_0$ is the background refractive index, and $\Delta n$ is the refractive index change that defines the sHCL. $H_0$ in Eq. (1) is the continuum Hamilton for wave propagation in the photonic lattice. If only the nearest-neighbor coupling of the waveguides is considered, the corresponding tight-binding Hamiltonian in the *k*-space can be written as[33]

$$H_T = t\begin{bmatrix} 0 & 0 & H_{13} & H_{14} & H_{15} \\ 0 & 0 & H_{13}^* & H_{14}^* & H_{15}^* \\ H_{13}^* & H_{13} & 0 & 0 & 0 \\ H_{14}^* & H_{14} & 0 & 0 & 0 \\ H_{15}^* & H_{15} & 0 & 0 & 0 \end{bmatrix}, \quad (2)$$

where $H_{13} = \exp(-ik_y a)$, $H_{14} = \exp[i(\sqrt{3}k_x a/2 + k_y a/2)]$, $H_{15} = \exp[i(-\sqrt{3}k_x a/2 + k_y a/2)]$, $a$ is the lattice constant, $t$ is the nearest neighbor coupling constant, and $H_{mn}^*$ denotes the complex conjugate of $H_{mn}$. Diagonalizing $H_T$ yields the band structure $\beta(\mathbf{k})$ as shown in Fig. 1(b). It consists of five bands, including a completely flat band touching two dispersive conical bands at the center (Γ point) of the first BZ, which resembles the pseudospin-1 Dirac cone in the Lieb lattice. Interestingly, there are also additional graphene-like pseudospin-1/2 Dirac cones at the BZ corners. In this work, we focus on the flatband state, and the corresponding Bloch eigenstate of the flat band is $|\beta,\mathbf{k}\rangle = [0, 0, -\sin(\sqrt{3}k_x), \sin(\sqrt{3}k_x/2 - 3k_y/2), \sin(\sqrt{3}k_x/2 + 3k_y/2)]^T$, which indicates that the flatband states have zero amplitudes on A and B sites. The fundamental CLS corresponds to filling the A, B sites with zero amplitude and other sites with equal amplitude but alternating opposite phase in each hexagonal plaquette [see red dashed square in Fig. 1(a)], and it is localized due to destructive interference. Typical experimental result of the intensity pattern of such a CLS is shown in the inset of Fig. 1(a). However, due to the flatband touching with the dispersive bands, there should be flatband line states localized in one direction but extended in the orthogonal direction to complete the flatband basis in the sHCL. As in the Lieb lattice[29], these line states can exist in a finite lattice with specially tailored boundaries satisfying the requirement of destructive interference. Considering the lattice geometry of the sHCL in real space and its band touching in *k*-space, we propose two types of line states (straight and zigzag) as illustrated in Figs. 1(c) and 1(d). The straight line state is along *x*-direction [Fig. 1(c)] while the zigzag one is along *y*-direction [Fig. 1(d)], both of them exist in a finite sHCL terminated with D and E sites.

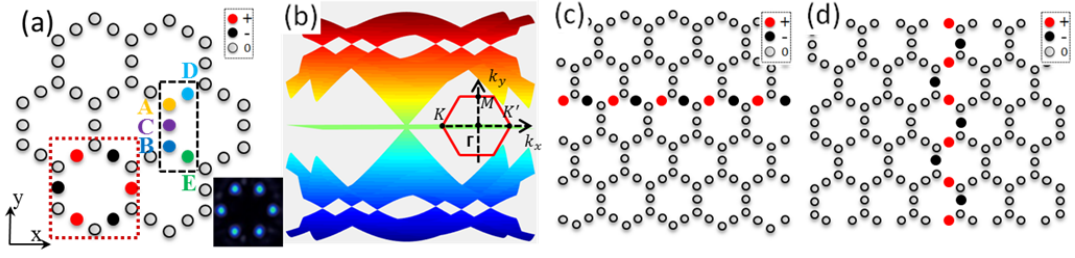

**Fig. 1. Illustration of flatband line states in a photonic sHCL**. (a) Schematic of sHCL structure consisting of five sites (A,B,C,D,E) per unit cell shown in dark-dashed square, with a fundamental flatband mode (the CLS) shown in red-dashed square. The right-bottom inset shows the experimentally obtained intensity pattern of the CLS. (b) Calculated band structure based on the tight-binding model showing a Lieb-like (pseudospin-1) Dirac cone intersected by a flat band at the first BZ center, and six graphene-like (pseudospin-1/2) Dirac cones located at the corners of the first BZ. The red hexagon depicts the first BZ marked with high symmetry points. (c, d) The sHCL under two different boundary cuttings that support (c) a "straight" line along $x$-direction and (d) a "zigzag" line along $y$-direction. For all figures, sites with zero amplitudes are denoted by gray color, and those with nonzero amplitudes of opposite phase are denoted by red and black colors.

In experiment, the sophisticated sHCL cannot be readily created by simple optical induction technique based on multi-beam interference. As such, we employ the cw-laser writing technique to establish such a lattice with desired lattice boundaries by site-to-site waveguide "writing" in a non-instantaneous nonlinear SBN crystal. The experimental setup is quite similar to that used in our recent work of flatband line states in photonic Lieb lattices[29]. Figure 2(a) shows a sHCL with edge termination on D and E sites along $y$-direction corresponding to Fig. 1(c), which supports the straight line states. The lattice spacing is about 32 $\mu$m, and a single-site excitation of a Gaussian beam leads to discrete diffraction with energy mainly coupling to the nearest-neighbor waveguides after 10-mm propagation through the crystal [see inset in Fig. 2(a)], which justifies that the waveguide coupling in the lattice satisfies the tight-binding condition. To observe the straight line states displayed in Fig. 1(c), the probe beam is shaped into a horizontal stripe pattern, with its phase modulated by a spatial light modulator (SLM) to make sure that the adjacent spots have opposite phase [Fig. 2(b1)]. Without the lattice, the input beam profile is not preserved after free-space propagation [Fig. 2(b2)]. In contrast, with the sHCL, its overall intensity pattern is well maintained [Fig. 2(b3)]. Furthermore, each spot remains localized and the out-of-phase feature persists as verified by the measured interferogram [Fig. 2(b5)]. For comparison, corresponding results for an in-phase stripe are presented in Figs. 2(c1-c3). In this case, the input beam cannot be confined, as the energy couples to zero-amplitude lattice sites A and B [Fig. 2(c3)], although the in-phase feature does not change [Fig. 2(c5)]. Due to limited propagation distance in experiment (1 cm), we perform numerical simulations to further corroborate our experimental observations. By taking same parameters used in our experiments but for a much longer propagation distance (4 cm), numerical results based on Eq. (1) are presented in [Figs. 2(b4) and 2(c4)]. One can see clearly the difference from output intensity patterns between the out-of-phase and in-phase cases.

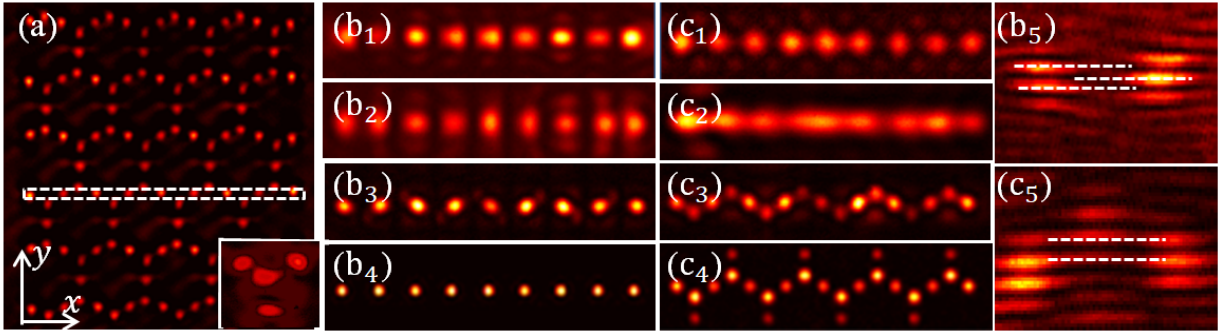

**Fig. 2. Experimental demonstration of photonic sHCL and a flatband straight line state**. (a) The laser-written sHCL, where the white dashed line marks the position of the probe beam and the lower inset shows discrete diffraction from single-site (site A) excitation. (b) From top to bottom: input out-of-phase probe beam, output without the lattice, output through the lattice, and simulation of (b3) for a longer propagation distance. (c) Corresponding results for an in-phase probe beam. (b5, c5) Zoomed-in interferograms of (b3, c3), respectively.

Next, we present the experimental results of zigzag line states extended along $y$-direction, for which the sHCL edge truncation along $x$-direction is shown in Fig. 3(a). To excite the zigzag line state, the probe beam is shaped into the zigzag pattern and launched into the lattice vertically [see Fig. 3(a), in which the white dashed rectangle marks the input position]. One clearly finds that only the out-of-phase beam can evolve into a flatband zigzag line state, with localized intensity pattern [Fig.3(b3)] and out-of-phase structure [Fig. 3(b5)] after exiting the lattice. The in-phase beam is not preserved during propagation and its energy couples to the nearest-neighbor minority (A, B) sites [Figs. 3(c3) and 3(c5)]. The difference is much more evident from long propagation distances, as indicated by the simulations in Figs. 3(b4) and 3(c4). It is worth noting that the zigzag line states realized here do not exist in either the Lieb or the Kagome lattices. In fact, to our knowledge, such flatband zigzag line states have never been theoretically predicted before.

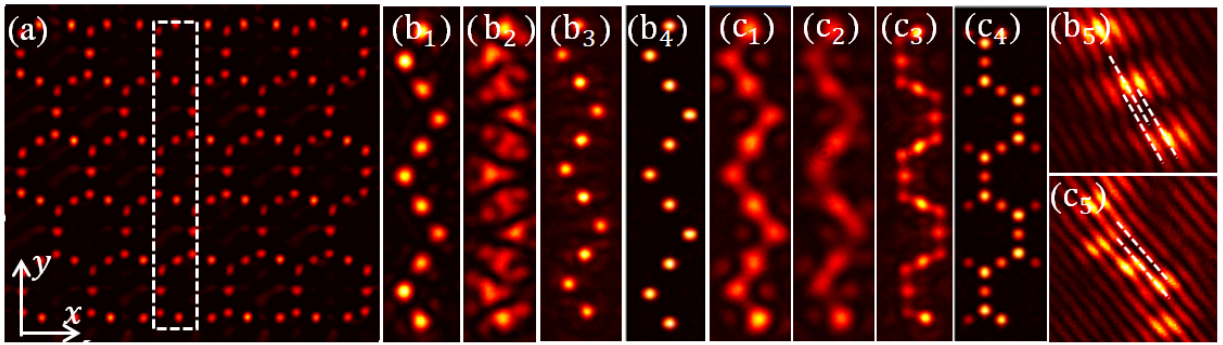

**Fig. 3. Experimental demonstration of photonic sHCL and a flatband zigzag line state**. Figure caption is the same as that for Fig. 2, except that the lattice boundaries of interest is are at the top and bottom, and the line state has a zigzag shape oriented vertically in y-direction.

Since the flatband line states are related to the flatband touching in $k$ space, it is instructive to experimentally measure the Fourier spectra of the line states. Typical experimental results are displayed in Fig.4, where the white dashed lines mark the edges of the 1st and 2nd BZs. From the band structure as shown in Fig. 1(b), one knows that the flatband touching in sHCL occurs at the 1st BZ center. Interestingly, the experimentally measured spectra of both the zigzag and straight line states mainly concentrate on higher-order BZ and do not touch the BZ center; the latter follows from the out-of-phase structure of the line states (with zero mean amplitude). For the straight line states the spectra mainly distribute in the 3rd BZ and touch four Γ points in the 2nd BZ center [Fig.4(a)], while the spectra of the zigzag line states cover all six Γ points in the 2nd BZ, and also additional modes locate at higher-order BZ centers [Fig.4(b)]. The experimental results of the spectra agree well with the simulation results [Figs. 4(c) and 4(d)]. Note that the spectra of the line states are in sharp contrast with those in the Lieb photonic lattices, where the spectra mainly distribute along the 1st BZ edge[29]. These results clearly show that the line states in sHCL are related to the flatband touching in $k$-space where the touching occurs at the BZ centers.

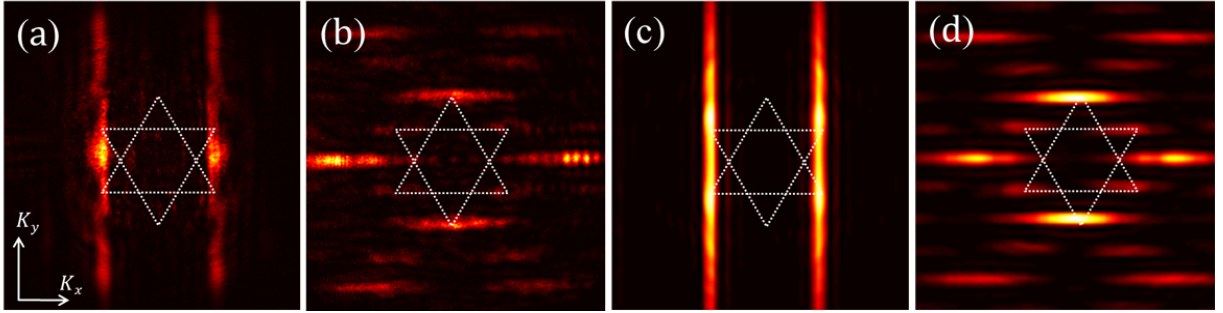

**Fig. 4. Momentum-space spectra of line states formed in sHCL**. (a, b) Experimentally measured and (c, d) numerically simulated $k$-space spectra for the (a, c) straight and (b, d) zigzag line states. White dashed lines outline the 1st and 2nd BZs.

In order to better understand the difference between the straight and zigzag line states, we first obtain their solutions based on the tight-binding model. If the sHCL is terminated along $x$ direction, then $k_x$ is not a good quantum number, so the Bloch eigenstate can be written as: $|\beta\rangle = [0,0,0,-1,+1]^T$. Clearly, the amplitude is zero on sublattices A, B and C, and nonzero on sublattices D and E with opposite phase. The analytical result completely agrees with the straight line state shown in Fig. 1(c). Similarly, if the sHCL lattice is ended along $y$-direction, the Bloch eigenstate reduces to $|\beta,k_x\rangle = [0,0,-2\cos(\sqrt{3}ak_x/2),+1,+1]^T$. Now the line state occupies C, D and E sublattices, different from the straight line state that only occupies two sublattices. If the probe is launched into the lattice at normal incidence $k_x = 0$, the Bloch eigenstate is recast into $|\beta\rangle = [0,0,-2,+1,+1]^T$. One finds that the line state has nonzero amplitude only on C, D and E sublattices, and the site C is out of phase with sites D and E. In fact, this is the unusual characteristics of the zigzag line state, along with that the amplitude on site C is two times larger than that on sites D and E. The reason can be quite intuitive: there are infinite numbers of independent zigzag line states in the horizontally cut sHCL if it is infinitely long along the $x$-direction and ended on sites D and E; however, every nearest two zigzag line states share and only share the same site C, demanding that the amplitude on site C be two times larger.

Another intriguing question is why the zigzag line states occupy three majority sites (C, D, E) but the straight line states only occupy two sites (D, E). Bear in mind that the sHCL does not change if one rotates it 60° clockwise or anti-clockwise due to lattice symmetry. Therefore, in addition to the straight line states along *x*-direction, there are also two additional line states along other two translation vector directions with an angle of 120° or 60° relative to the positive *x* axis, which can be described by $|\beta\rangle = [0,0,-1,+1,0]^T$ (occupying sites C and D), or $|\beta\rangle = [0,0,-1,0,+1]^T$, (occupying sites C and E). Appropriate linear superposition of the two straight line states not only gives rise to $|\beta\rangle = [0,0,-2,+1,+1]^T$, but also $|\beta\rangle = [0,0,0,-1,+1]^T$. The former one is exactly the solution which describes the zigzag line state, while the latter one is the straight line state. In other words, the zigzag line state is also a reflection of the straight line state, and its appearance is dependent on the spatial geometry of the finite sHCL.

To further confirm the existence of the above observed straight and zigzag flatband line states in finite sHCLs, we numerically solve Eq. (1) directly by considering the finite sHCL boundary conditions properly and taking parameters similar to those used in experiments. We seek for the solution $\Psi(x,y,z) = u(x,y)\exp(i\epsilon z + ik_\mu \mu)$ with $\epsilon$ being the propagation constant, $u(x,y)$ the envelope of line states, and $\mu$ being $y$ ($x$) for the straight (zigzag) line state. Plugging this solution ansatz into Eq. (1), one obtains an eigenvalue problem, based on which the band structure of the truncated semi-infinite sHCL can be obtained by using the plane-wave expansion method. For convenience, we introduce the dimensionless propagation constant $\beta = k_0 r_0^2 \epsilon$ with $r_0$ being the real probe beam width.

In Fig. 5(a), the band structure of the truncated sHCL (one unit cell) shown in Fig. 5(b1) is displayed. In the band structure, the flat band is quite obvious and the dispersive bands are almost symmetric about the flat band. Therefore, the band structure obtained based on the continuum model can be also utilized to demonstrate the validity of our previous discussions based on the tight-binding model. We choose the flattest case in the flat band and mark it with red color in Fig. 5(a), on which the amplitude and phase of the flatband mode with $k_y = 0$ are presented in Figs. 5(b2) and 5(b3), respectively. One finds that there are two independent straight line states extended along *x* direction in Fig. 5(b2), and every two nearest-neighbor sites (i.e., D and E sites) are out-of-phase according to the phase distribution in Fig. 5(b3). The numerical results obtained based on the continuum model are completely in accordance with the experimental results in Fig. 2 and the analytical results based on the tight-binding model.

Corresponding to the truncated sHCL (one unit cell) in Fig. 5(d1), the band structure is displayed in Fig. 5(c). Similar to Fig. 5(a), we also choose one flattest case and mark with red color in Fig. 5(c). The flatband mode with $k_x = 0$ on this flattest band is exhibited in Fig. 5(d2), and its corresponding phase is shown in Fig. 5(d3). There are also two independent zigzag line states in one unite cell. Again, one finds that the results agree with the analytical results based on the tight-binding model: there is a π phase shift between the amplitudes on sites C, D and E, and the amplitude on site C is two times larger than that on sites D and E.

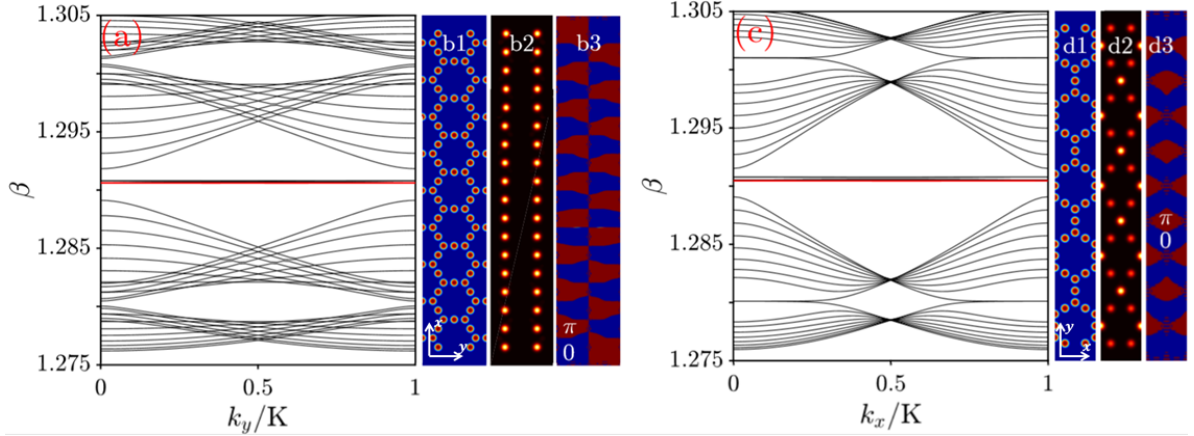

**Fig. 5. Numerical results of line states in sHCL based on continuum model.** (a, b) The straight line and (c, d) zigzag line state solutions, where (a, c) are band structure with K representing the BZ width, (b1, d1) the unit cell of the lattice structure, (b2, d2) the amplitudes of the flatband modes with $\beta$ indicated by the red lines in (a, c) and $k_\mu = 0$, and (b3, d3)the corresponding phase of (b2, d2). Note that the amplitudes are not equal for all sites in (d2) and there are phase differences as indicated in the corresponding areas.

In conclusion, we have experimentally realized finite-sized photonic sHCLs with desired boundaries by employing the cw-laser writing method in a nonlinear bulk crystal, and more importantly, demonstrated for the first time straight and zigzag flatband line states manifesting the NLSs in infinite lattices. The two line states occupy different sublattice sites in real space, and exhibit different spectra in momentum space. Interestingly, the zigzag line state can be viewed as a linear superposition of straight line states along different directions. Moreover, the spectra of the line states in momentum space distribute in higher-order BZ and occupy the BZ center where the flatband touches the dispersive bands, reflecting unusual characteristic of line states arising from band-touching. Our work not only reveals new flatband states, but also brings about new possibilities to explore both flatband and Dirac physics in one platform, as the sHCL represents an ideal system where fermionic (spin-1/2) and bosonic (spin-1) Dirac points coexist[31,32].

**Acknowledgement**: This work is supported by the National Key R&D Program of China (2017YFA0303800), the Chinese National Science Foundation (91750204, 11674180 and 11922408), PCSIRT (IRT_13R29), and 111 Project (No. B07013) in China.